\documentclass[twocolumn,prb]{revtex4-1}

\usepackage[english]{babel}

\usepackage{graphicx}
\usepackage{amsmath}
\usepackage{amsfonts}
\usepackage{amssymb}
\usepackage{bm,bbm}
\usepackage{color}

\usepackage{hyperref}

\newcommand{\tr}{\mathrm{tr}}

\def\tit#1{{\em #1},}
\def\etal#1{ {\em et al.}}

\begin{document}

\title{Absence of superdiffusion in the quasiperiodic spin chain at weak integrability breaking}

\author{Marko \v Znidari\v c}
\affiliation{Department of Physics, Faculty of Mathematics and Physics, University of Ljubljana, Jadranska 19, SI-1000 Ljubljana, Slovenia}

\date{\today}

\begin{abstract}
  There has been interest in the spin transport properties of the Aubry-Andr\' e-Harper model at high temperatures under weak integrability breaking, in particular for small interactions or small fields. We present old unpublished and new results that show that the model is diffusive, meaning that the claimed superdiffusion~\cite{EPL17,swingle20} is a finite-size effect.
\end{abstract}

\maketitle

The model we consider is a one-dimensional chain of spin-1/2 particles called the Aubry-Andr\' e-Harper (AAH) model described by the Hamiltonian
\begin{equation}
  H=\sum_{j=1}^{L-1} \sigma_j^{\rm x}\sigma_{j+1}^{\rm x}+\sigma_j^{\rm y}\sigma_{j+1}^{\rm y}+\Delta \sigma_j^{\rm z}\sigma_{j+1}^{\rm z}+\lambda(h_j\, \sigma_j^{\rm z}+h_{j+1} \sigma_{j+1}^{\rm z}),
\end{equation}
with $h_j=\cos{(2\pi \beta j)}$ and $\beta=\frac{\sqrt{5}-1}{2}$. In the fermionic language $\Delta$ is the interaction strength. The model is of high interest to theoretical and experimental physicists (as well as mathematicians) for its interesting properties, being due to competition between hopping (ballistic transport), quasiperiodic potential (marginal localization), and interaction, for details consult e.g. refs. in~\onlinecite{EPL17,PNAS18,swingle20}.

We are focusing on spin transport at infinite temperature, i.e., for generic initial states. The model is integrable at two points. For $\Delta=0$ one has noninteracting fermions, with ballistic spin transport for $\lambda<1$ and localization for $\lambda>1$. When $\lambda=0$ one has again an integrable model, but this time of the nontrivial Bethe-ansatz type, where one has ballistic spin transport for $|\Delta|<1$. We want to address spin transport at small integrability breaking, that is at small $\Delta$ or small $\lambda$.

Spin transport in this regime has been studied for instance in Ref.~\onlinecite{EPL17}, where it has been claimed that (i) the model displays superdiffusion for small interactions $\Delta$ and $\lambda=0.75$, as well as for small potential amplitudes $\lambda$ at $\Delta=0.5$, and (ii) that there is no finite region of parameters with diffusive transport. More recent Ref.~\onlinecite{swingle20} also claims that (iii) one has superdiffusion at small $\lambda$ and $\Delta=1$. We show that all these statements are wrong. 

The problem is that correctly numerically assessing transport at small integrability breaking, i.e., when scattering between ballistic integrable ``modes'' is weak, is rather tricky. In principle things are elementary -- the Fermi's golden rule predicts that the scattering will go to zero as the perturbation strength decreases and therefore the corresponding scattering length increases. If one wants to have any chance of observing the correct asymptotic behavior the system's length $L$ has to be much larger than the scattering length. In a similar model (one with an independent random potential rather than the quasiperiodic) it has been found empirically~\cite{PRL16} that in practice one can expect to need systems (much) larger than $\sim 100$ sites at say $\lambda \approx 0.1$. Provided these points are properly taken into account one expects to see diffusion in a nonintegrable system~\cite{diff} and not superdiffusion, which is indeed what has been observed~\cite{PNAS18} also in the AAH model at small $\Delta$. Here we focus on small $\lambda$ and $\Delta=1$ studied in Ref.~\onlinecite{swingle20}.
\begin{figure}[t!]
  \centerline{\includegraphics[width=2.6in]{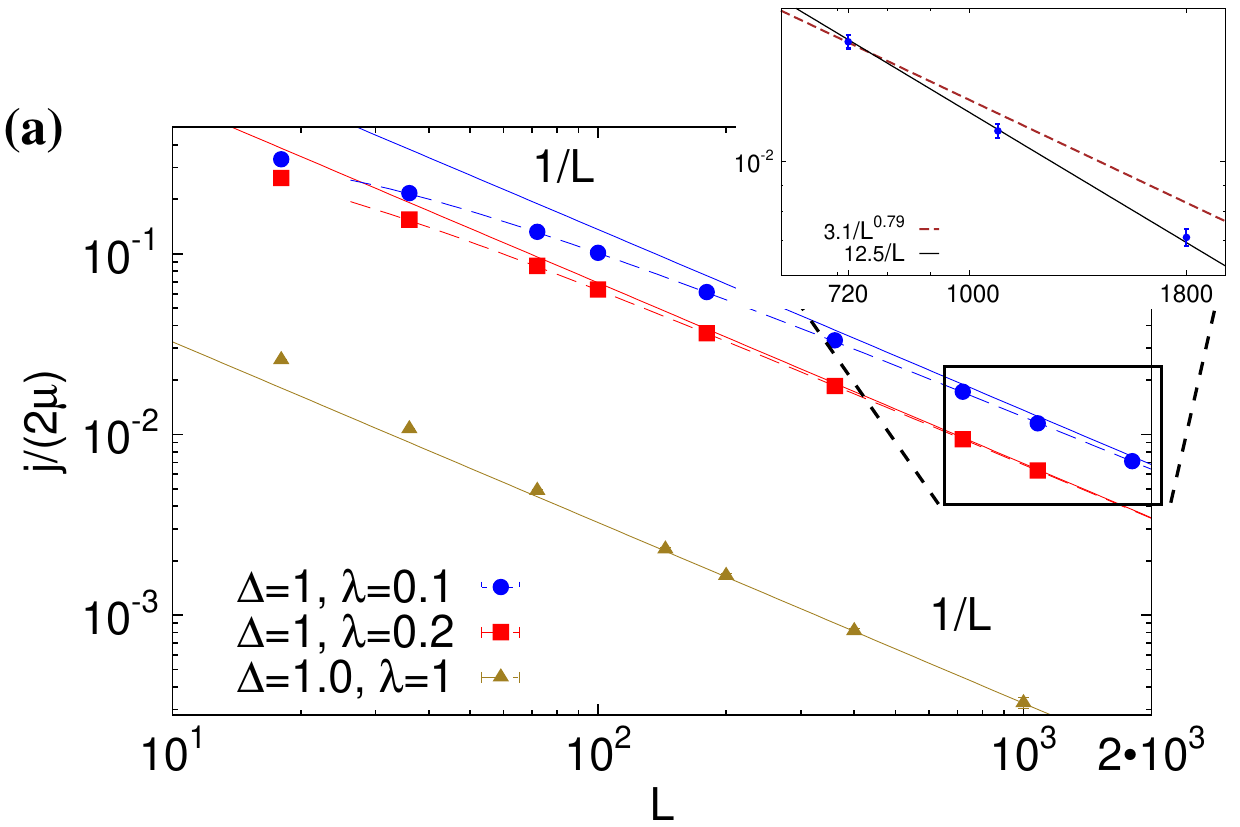}}
  \centerline{\includegraphics[width=2.5in]{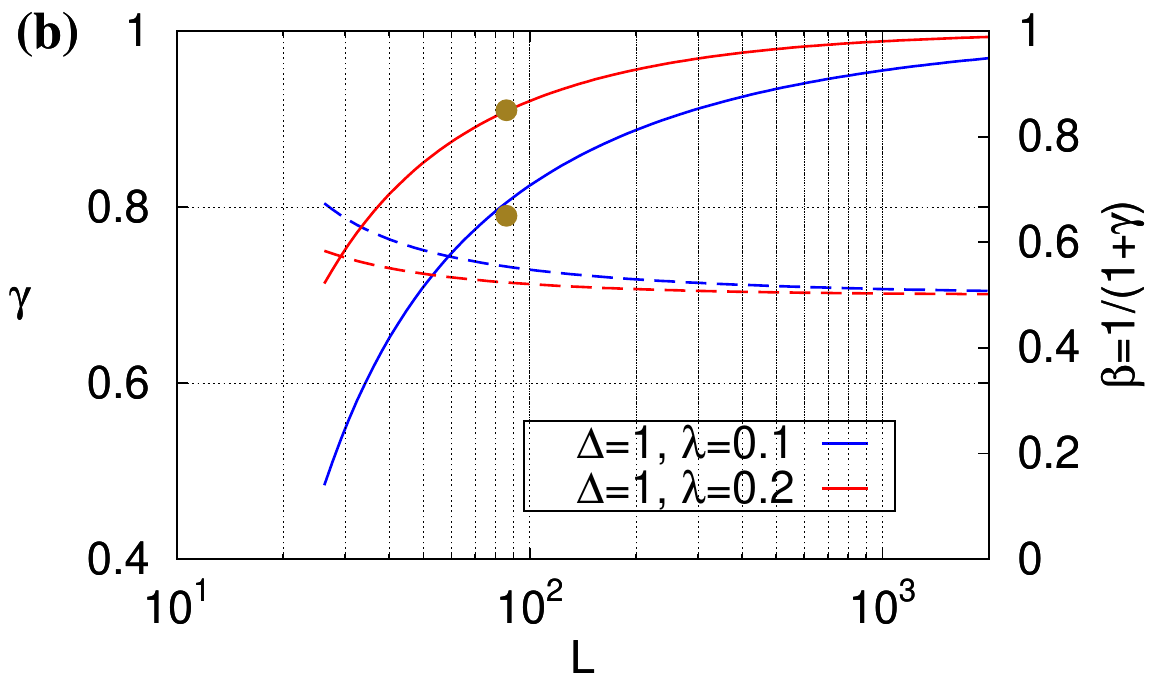}}
  \caption{(a) Diffusive transport in the strongly interacting AAH model (NESS current scaling as $j \sim 1/L$). The inset shows the largest three $L$, where we can see that $\sim 1/L^{0.79}$ is by several sigma away from the data. (b) Dashed curves in (a) are fits of the form $j \sim 1/L(1-b/L^\alpha)$ with $b$ a fitting parameter and $\alpha\approx 0.5-0.8$, and are used in (b) to plot a finite-size $\gamma$ (full curves), i.e., the local slope in the log-log plot of $j(L)$. Full circles are data points from Fig.4 of Ref.~\onlinecite{swingle20}, for the explanation of $\beta$ and dashed curves see Fig.~\ref{fig2}.}
\label{fig1}
\end{figure}

\begin{figure}[t!]
  \centerline{\includegraphics[width=2.5in]{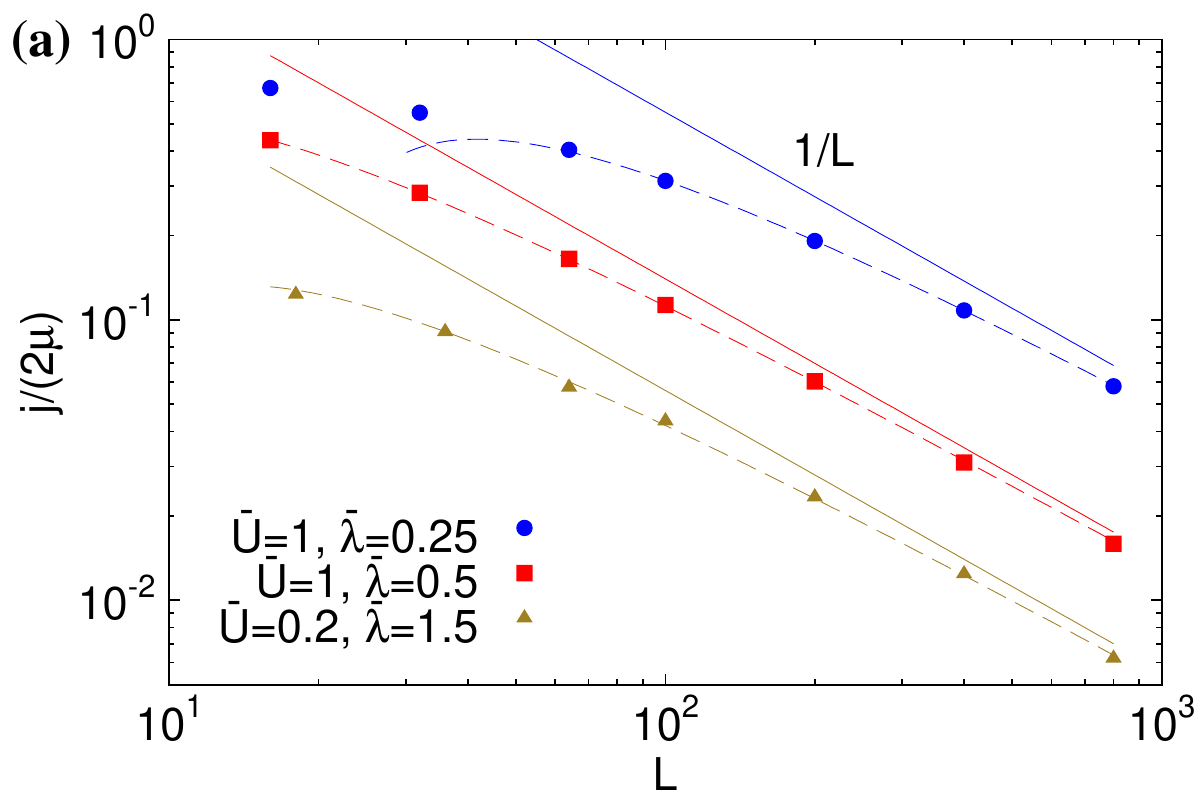}}
  \centerline{\includegraphics[width=2.5in]{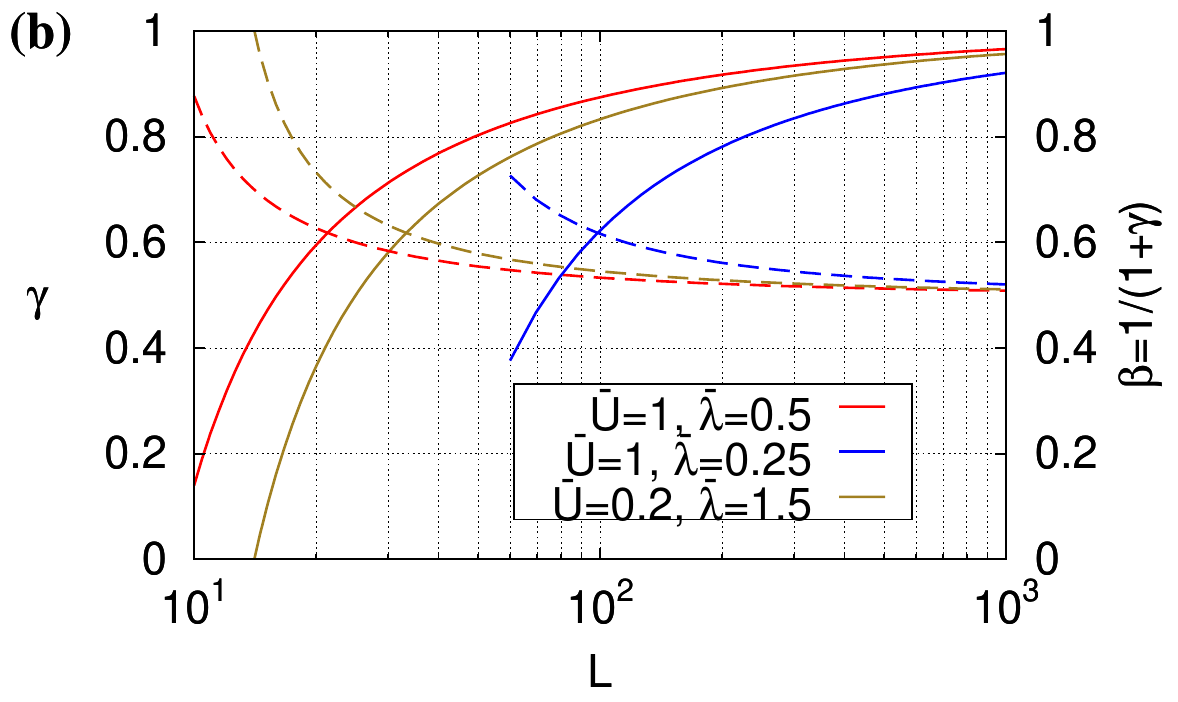}}
  \caption{(a) Diffusive asymptotic scaling of the NESS current (full lines) in the AAH model. Reported parameters $\bar{U}$ and $\bar{\lambda}$ are the same as $U$ and $\lambda$ in Ref.~\onlinecite{EPL17}, and are $\bar{U}=2\Delta$ and $\bar{\lambda}=2\lambda$. (b) The corresponding finite-size scaling exponent $\gamma$ (left axis, full curves) and the unitary spreading exponent $\beta=1/(1+\gamma)$ (right axis, dashed curves) is plotted from the dashed fits in (a). Ref.~\onlinecite{EPL17} reports superdiffusive $\beta\approx 0.70, 0.85$ at the shown parameters (red and blue, respectively).}
  \label{fig2}
\end{figure}
To be able to study large systems we use the Lindblad master equation, see e.g.~\onlinecite{PRL16} and Eq.~(S1) in Ref.~\onlinecite{PNAS18} for units and conventions that we use. Exactly the same method has been used in Ref.~\onlinecite{swingle20}, the only difference here is that we study more than $10\times$ larger systems. Briefly, the system is driven at its boundaries by magnetization driving with a (chemical) potential difference $2\mu$ ($\mu\ll 1$). After long time the system reaches a nonequilibrium steady state $\rho_\infty$ (NESS) in which one has a nonzero magnetization gradient (at the edges one has $\tr{(\rho_\infty \sigma_{1,L}^{\rm z})} \approx \pm \mu$) and a nonzero spin current $j$ flowing through the system, $j=\tr{(\rho_\infty [2\sigma_i^{\rm x}\sigma_{i+1}^{\rm y}-2\sigma_i^{\rm y}\sigma_{i+1}^{\rm x}])}$. By studying the scaling of the current with $L$, $j \sim 1/L^\gamma$, one gets the transport type from the power $\gamma$. If $\gamma=1$ one has diffusion, if $\gamma<1$ one has superdiffusion. Ref.~\onlinecite{swingle20} used a power-law fitting in a tiny window $L \in [72,100]$ to extract~\cite{foot2} $\gamma$, and this lead to incorrect conclusions.

Because calculations are very time-consuming~\cite{foot1} we picked just two values of $\lambda$. For $\lambda=0.1$ we estimated that it will be the easiest to demonstrate that there is no superdiffusion (the claimed~\cite{swingle20} superdiffusive exponent is $\gamma \approx 0.79$ and is sufficiently away from $1$). The second value $\lambda=0.2$ is used to demonstrate that there is a whole diffusive phase, not just a single point. Our data is summarized in Fig.~\ref{fig1}. The results show that one has diffusion (superdiffusion deviates from the data by several sigma -- the inset to Fig.~\ref{fig1}(a)), and that one needs system sizes well in excess of $L \sim 100$ to observe it (Fig.~\ref{fig1}(b)). 

The method becomes less efficient~\cite{JSTAT20} at large $\lambda$, where Ref.~\onlinecite{swingle20} reports subdiffusion (though always with $\gamma$ being very close to diffusive $1$); for instance, at $\lambda=1$ they get $\gamma \approx 1.05$. Our data in Fig.~\ref{fig1}(a) is at $\lambda=1$ instead compatible with diffusion~\cite{foot3}. We conservatively estimated the error in our $j(L=1000)$ to be $\approx 7\%$; thus, requiring the agreement with $1/L^\gamma$ for $L>100$ (which we do get with $\gamma=1$) we can estimate that with $\gamma=1.05$ the deviation at $L=1000$ would be about $(1000/100)^{0.05}\approx 12\%$, which is more than our estimated error. We therefore conclude that at $\lambda=1$ one has $\gamma \approx 1.00 \pm 0.03$, and is unlikely that $\gamma \ge 1.05$. 

Finally, we briefly touch upon Ref.~\onlinecite{EPL17} where $\Delta=0.5$ is used. Doing the same analysis as before we get data in Fig.~\ref{fig2}, again showing diffusion with no signs of either super- or subdiffusion. The observed~\cite{EPL17} superdiffusion is therefore a finite-time effect (displacement variance was fitted with a power-law in a window $t \in [2,10]$) and will disappear at longer times. Using different method that can go to larger times~\cite{EPL17} does not alleviate the problem because the used perturbation there is even smaller.



\begin{thebibliography}{}
\bibitem{foot2} It is hard to understand why 4 (redundant) L were calculated in that small interval rather than say an additional $L=200$ where one would already get hints that something is amiss. For a power-law fitting equidistant points in a log-scale matter.
\bibitem{foot3} Ref.~\onlinecite{PNAS18} also observed only diffusion in the AAH model (at different parameters than here). In Ref.~\onlinecite{swingle20} it is incorrectly stated that~\onlinecite{PNAS18} observed an "intermediate slow dynamics''.
\end{thebibliography}
\end{document}